\documentclass[conference]{IEEEtran}

\usepackage{graphicx}
\usepackage{times}
\usepackage{helvet}
\usepackage{courier}
\usepackage{balance}

\usepackage{multirow}
\usepackage{amsmath}
\usepackage[font=footnotesize]{subfig}
\usepackage{fixltx2e}
\usepackage{url}
\usepackage{epstopdf}
\usepackage{amsthm}
\usepackage{color,soul,xspace}
\usepackage{multirow}

\newcommand{\vsa}{\vspace*{-0.35cm}}

\begin{document}

\title{The Social Media Genome: Modeling Individual Topic-Specific Behavior in Social Media}

%\author{Petko Bogdanov$^\dagger$, {\bf Michael Busch$^\dagger$,} \\ {\bf \Large Jeff Moehlis, Ambuj K. Singh} \\
%University of California Santa Barbara, \\
%Santa Barbara, CA 93106 \\
%\And
%Boleslaw K. Szymanski \\
%Rensselaer Polytechnic Institute\\
%110 8th St., Troy NY 12180
%}

\author{
  \IEEEauthorblockN{Petko Bogdanov$^\dagger$, Michael Busch$^\dagger$, Jeff Moehlis, Ambuj K. Singh}
  \IEEEauthorblockA{University of California Santa Barbara\\
		    Santa Barbara, CA 93106 }
  \and
  \IEEEauthorblockN{Boleslaw K. Szymanski}
  \IEEEauthorblockA{Rensselaer Polytechnic Institute\\
	            110 8th St., Troy NY 12180}
}

\maketitle
{\let\thefootnote\relax\footnotetext{$^\dagger$These authors contributed equally to this work. Corresponding authors, petko@cs.ucsb.edu, mbusch@engr.ucsb.edu.}}

\begin{abstract}
%\begin{quote}
Information propagation in social media depends not only on the static follower structure but also on the topic-specific user behavior. Hence novel models incorporating dynamic user behavior are needed.  
To this end, we propose a model for individual social media users, termed a \textit{genotype}. The genotype is a \textit{per-topic} summary of a user's interest, activity and susceptibility to adopt new information. We demonstrate that user genotypes remain invariant within a topic by adopting them for classification of new information spread in large-scale real networks. Furthermore, we extract topic-specific influence backbone structures based on information adoption and show that they differ significantly from the static follower network. When employed for influence prediction of new content spread, our genotype model and influence backbones enable more than $20\%$ improvement, compared to purely structural features. We also demonstrate that knowledge of user genotypes and influence backbones allow for the design of effective strategies for latency minimization of topic-specific information spread.
\end{abstract}

\section{Introduction}
Trends and influence in social media are mediated by the individual behavior of users and organizations embedded in a follower/subscription network. The social media network structure differs from a friendship network in that users are allowed to \textit{follow} any other user and follower links are not necessarily bi-directional. While a link enables a possible influence channel, it is not always an active entity, since a follower is not necessarily interested in all of the content that a \textit{followee} posts. Furthermore, two individuals are likely to regard the same token of information differently. Understanding how information spreads and which links are active requires characterizing the users' individual behavior, and thus going beyond the static network structure. A natural question then arises: \textit{Are social media users consistent in their interest and susceptibility to certain topics?}

In this work, we answer the above question by demonstrating a persistent topic-specific behavior in real-world social media. We propose a user model, termed \textit{genotype}, that summarizes a user's topic-specific footprint in the information dissemination process, based on empirical data. The social media genotype, similar to a biological genotype, captures unique user traits and variations in different genes (topics). Within the genotype model, a node becomes an individual represented by a set of unique invariant properties.

For our particular analysis, the genotypes summarize the propensity and activity level in adoption, transformation, and propagation of information within the context of different topics. We propose a specific set of properties based on adopting topic-specific Twitter hashtags---tokens that annotate messages and allow users to participate in global discussions~\cite{Tsur2012}. The model, however, applies to more general settings capturing, for example, dissemination of urls or sentiment-charged messages. % As to be discussed in the \textit{Genotype Model and Construction of topic genotypes} section, the genotype can be viewed as a framework, extensible to other measures of topical behavior.
%Next, we analyze the topic-specific influence network based on adoption of new information and demonstrate that its structure differs significantly from the overall follower network. Our analysis provides empirical evidence that follower links are not necessarily associated with influence and that not all of them are essential in information spread and trends formation. Furthermore, we show that users tend to react to and adopt certain information along topic-specific influence backbones and their behavior remains relatively consistent for all information items within a topic. 

We construct the genome (collection of user genotypes) of a large social media dataset from Twitter, comprised of both follower structure and associated posts. The observation of stable genotypes (behavior) leads to natural further questions: \textit{Can the consistent user behavior be employed to categorize novel information based on its spread pattern? Can one utilize the genotypes and the topic-specific influence backbone to (i) predict likely adopters/influencers for new information from a known topic and (ii) improve the network utility by reducing latency of disseminated information?} We explore the potential of the genotype model to answer the above questions within the context of Twitter. 

To validate the consistency of genotypes, we show that combining genotype-based classifiers into a composite (network-wide) classifier achieves accuracy of $87\%$ in predicting the topic of unobserved hashtags that spread in the network. We extract and analyze topic-specific influence backbone structures and show that they differ from the static follower network. We, then, turn to two important applications: influence prediction and topic-specific latency minimization. We achieve $20\%$ improvement in predicting influencers/adopters for novel hashtags, based on our model, as compared to relying solely on the follower structure. We also demonstrate that knowledge of individual user genotypes allows for effective reduction in the average time for information dissemination (more than 40\% reduction by modifying the behavior of 1\% of the nodes).

Our contributions, in the order they are presented, include: \textit{(i)} proposing a genotype model for social media users' behavior that enables a rich-network analysis; \textit{(ii)} validating the consistency of the individual genotype model; \textit{(iii)} quantifying the differences of behavior-based influence backbones from the static network structure in a large real-world network; and \textit{(iv)} employing genotypes and backbone structure for adopter/influencer prediction and latency minimization of information spread.

\section{Related Work}
The network structure has been central in studying influence and information dissemination in traditional social network research~\cite{Kempe2003,Kimura2010}. Compared to them, large social media systems exhibit a relatively denser follower structure, non-homogeneous participation of nodes, and topic specialization/interest of individual users. Twitter, for example, is known to be structurally different from human social networks~\cite{Kwak2010}, and the intrinsic topics of circulated hashtags are central to their adoption~\cite{Romero2011}. %Nodes' affinity to topics play an even more central role in formation of new trends/vocabulary and success of memes. %Since these phenomena are carried out by social agents, a relevant question is \textit{how can one characterize the transmission/modification behavior or bias of social agents in a network?}

A diverse body of research has been dedicated to understanding influence and information spread on networks, from theories in sociology~\cite{Friedkin2006} and epidemiology~\cite{Newman2003}, leading to empirical \textit{large-scale} studies enabled by social web systems~\cite{Romero2011,Dodds2011,Yang2011,Bandari12}. %The majority of this research focuses on the importance of the follower structure among nodes, assuming that all links are active in information spreading. In contrast, 
Here, we postulate that the influence structure varies across topics~\cite{Weng2010} and is further personalized for individual node pairs. Lin and colleagues~\cite{Lin2011} also focus on topic-specific diffusion by co-learning latent topics and their evolution in online communities. The diffusion that the authors of \cite{Lin2011} predict is implicit, meaning that nodes are part of the diffusion if they use language corresponding to the latent topics. In contrast, we focus on topic-specific user genotypes and influence structures concerned with passing of observable information tokens and their temporal adoption properties.  

Earlier data-centered studies have shown that sentiment~\cite{Dodds2011} and local network structure~\cite{Romero2011} have an effect on the spread of ideas. The novelty of our approach is the focus on content features to which users react. Previous content-based analyses of Tweets have adopted latent topic models~\cite{Suh2010,Ramage2010}. We tie both content and behavioral features to network individuals. 

% influence structure
With regards to influence network structure and authoritative sources discovery, Rodriguez and colleagues~\cite{Gomez2013} were able to infer the structure and dynamics of information (influence) pathways, based on the spread of memes or keywords. Bakshy et al.~\cite{Bakshy2011} focus on Twitter influencers who are roots of large cascades and have many followers, while Pal et al~\cite{Pal2011} adopt clustering and ranking based on structural and content characteristics to discover authoritative users. Although the above works are similar to ours in that they focus on influence structures and user summaries, our genotype targets capturing the invariant user behavior and information spread within topics as a whole, involving a collection of topically related information parcels.

\section{Genotype Model}
\begin{table*}[t]
\centering
{\scriptsize
\begin{tabular}{|l|p{.41\linewidth}|p{.44\linewidth}|	} \hline
Metric & Function definition & Notes \\\hline
\textit{Time} & $\textrm{TIME}(u,h)=min_{(u,h)}(t(u,h))-min_{v\in V_u}(t(v,h))$, where $t(u,h)$ is the time $(u,h)$ occurs and $V_u$ is the set of followees of $u$. & The absolute amount of time between a users first exposure to the given hashtag and his first use of that same hashtag.\\\hline
\textit{Number of Uses}& $\textrm{N-USES}(u,h)=\left|\{(u,h)\}\right|$, where $|\cdot |$ is the cardinality function. & The total number of occurrences of the $(u,h)$ pair.\\\hline
\textit{Number of Parents} & $\textrm{N-PAR}(u,h)=\left|\{v\in V_u\;|\;t(v,h)<t(u,h)\}\right|$ & The number of followees to adopt before the given user.\\\hline
\textit{Fraction of Parents} & $\textrm{F-PAR}(u,h)= \left|\{v\in V_u \;| \;t(v,h)<t(u,h)\}\right|/\left|V_u\right|$. & The fraction of a user's followees who have adopted the hashtag prior to the user.\\\hline
\textit{Latency} & $\textrm{LAT}(u,h)= \left(\left|\{h_j\in H_{T_i}\;|\; H_{T_i}\ni h, \textrm{ and } t(u,h_j)<t(u,h)\}\right|\right)^{-1}$. & The inverse of the number of same-topic hashtags posted to the user's time-line between his first exposure to the hashtag and his first use of the hashtag.\\\hline
\textit{Log-latency} & $\textrm{LOG-LAT}(u,h)=log\left(\textrm{LAT}(u,h)/ Avg(\textrm{LAT}(w,h)\right. \textrm{s.t.} \left. w\in U)\right)$. & The logarithm of each latency value after each latency value has been divided by the mean latency value for that hashtag.\\\hline
\end{tabular}	
}
\caption{Behavior-based metrics that are components of the topic-specific user genotype.}
\label{tab:metrics}
\vsa
\end{table*}
Here we define our genotype model that captures the topic-specific behavior of a single user (node) within a social media network. Our main premise is that, based on observed network behavior, we can derive a unique signature of a user. Hence, the genotype model is a user model, by definition, in the sense that it is an abstract representation of a social network user. For our analysis, this signature captures adoption and reposting of new information, activity levels, and latency of reaction to new information and influential neighbors. Although other behavior traits can be incorporated as well, we seek to summarize user behavior with respect to a set of predefined topics. %, since users naturally exhibit a different affinity to topics \todo{reference?}.

%% general abstract description of Genotype and follower network
A social media network $N(U,E)$ is a set of users (nodes) $U$ and a set of follow links $E$. A directed follow link $e=(u,v),e \in E$ is connects a source user $u$ (\textit{followee}) to a destination user $v$ (\textit{follower}). The network structure determines how users get exposed to information posted by their followees. The static network does not necessarily capture influence as users do not react to all information to which they are exposed. To account for the latter, we model the behavior of individual users within the follower network within their genotypes. 

In its most general form, a user's genotype $G_u$ is an entity embedded in a multi-dimensional feature space that summarizes the \textit{observable behavior} of user $u$ with respect to different topics. %More formally, the genotype $G_u\in\mathrm{R}^T$ captures a set of measurable quantities (e.g. activity level, latency of reaction etc.) with respect to a set of topics $T$. 
It is up to the practitioner to define the different dimensions of the topic feature space, and an observable behavior in the network locality of a node. Each genotype value can be viewed as an allele that the user introduces to the process of message propagation through a network. 

In our study, we focus on hashtag usage within Twitter, since hashtags are simple user-generated tokens that annotate tweets generated by either a social group or a specific social phenomenon, and are often ``learned'' from others on the social network~\cite{Tsur2012}. In this context, a hashtag serves as a genetic parcel of cultural information, just like alleles of a gene within a biological context. Hashtags can be associated with topics such that an individual's response to a collection of hashtags within a topic indicates a user's propensity to respond to other hashtags within that same topic. %Thus, in order to summarize a user's network behavior within a set of topics, we track their usage of related hashtags.

%As a concrete instantiation of our genotype model, we study the case when the elements of the user genotype $G_u$ are hashtag topic distributions within Twitter. 
We consider a set of predefined hashtags $H=\left\{h\right\}$, each associated with a topic $T_i\in T$. To obtain the genotype, we analyze the social media message (tweet) stream produced by a user $u$, with respect to $H$. Let us define $m(\cdot)$ to be a function that maps each occurrence of $(u,h)$ to a real values $m:\{(u,h)\}\mapsto \mathbf{R}$. Let $H_{(u,T_i)}:=\{h\}_{T_i} \cap \{h\}_u$, then the $i^{th}$ element of the user genotype $G_u$ is the set of $\{m(u,h)\; | \;h\in H_{(u,T_i)}\}$ values. We remark that this set of values may also be reduced to their average value or some approximated distribution function if one wishes to have a coarser representation of the data. %\todo{from now on replace mentions of ``genotype'' with ``each $G_u$''.}
To construct each user's topic-genotype from empirical data, we consider a variety of metrics $m(\cdot)$ for $(u,h)$ pairs, listed in Table~\ref{tab:metrics}.
These metrics serve the purpose of quantifying a user's response to a hashtag by defining the data values that are used to estimate the topic distributions. While TIME and N-USES are intuitively obvious metric choices, LAT and LOG-LAT are novel to this manuscript. N-PAR and F-PAR have been previously studied in a different context~\cite{Romero2011}, and are included here for comparison.

\section{Datasets}
We chose Twitter to analyze user behavior via our genotype model, since Twitter has millions of active users and messages have a known source, audience, time stamp, and content. Similar analysis can be performed in other social media networks with a known follower structure and knowledge of the shared content (memes, urls and buzz words) in time. 

\begin{table}[t]
\centering
{\scriptsize
\begin{tabular}{|l|c|c|c||c|c|c|} \hline
             & \multicolumn{3}{|c|}{SNAP (users=42M,tweets=467M)} & \multicolumn{3}{|c|}{CRAWL (users=9K,tweets=14.5M)} \\ \hline
 Topic       & Hashtags & Users & Uses/HT & Hashtags & Users & Uses/HT \\ \hline	
 Business    & 27       & 20k   & 1,155   & 19       & 1,493 & 88   \\ \hline 
 Celebrities & 32       & 26k   & 1,009   & -        & -     & -    \\ \hline
 Politics    & 485      & 349k  & 2,020   & 121      & 5,480 & 49   \\ \hline
 Sci/Tech    & 33       & 415k  & 6,889   & 63       & 4,982 & 100  \\ \hline
 Sports      & 98       & 76k   & 3,274   & 24       & 320   & 14   \\ \hline
\end{tabular}	
}
\caption{Statistics of the SNAP and CRAWL data sets.}
\label{tab:hastag_topic}
\vsa
\end{table}

\subsection{Twitter follower structure and messages}

We use two datasets from Twitter: a large dataset \textit{SNAP}~\cite{Yang2011} including a $20\%$ sample of all tweets over a six-month period and the complete follower structure~\cite{Kwak2010}; and a smaller CRAWL dataset containing all messages of included users that we collected using Twitter's public API in 2012, where we started from initial seed nodes (members of the authors' labs) and crawled the follower structure and related posts. SNAP includes a network-wide view for a 6 month period, while CRAWL provides longitudinal completeness for a smaller subnetwork of users. Statistics of the two datasets are summarized in Table~\ref{tab:hastag_topic} and further discussed in the Supplement~\cite{appendix}.

\subsection{Grouping hashtags into topics}
% why hashtags 
While hashtags present a concise vocabulary to annotate content, they are free-text user-defined entities. Hence, we need to group them into topics in order to summarize user behavior at the topic level. In this work we assume a 1:1 mapping of topics to hashtags, while in a more general framework disseminated hastags (urls, memes, etc.) can be ``softly'' assigned to more than one topic. We work with five general topics as dimensions for our user genotypes: Sports, Politics, Celebrities, Business and Science/Technology. We obtain a set of $100$ high-confidence hashtag annotations from a recent work by Romero and colleagues~\cite{Romero2011}, further augmented by a set of curated business-related hashtags~\cite{Ribiero2012}. %, obtained from an online list curated as recommendation for users interested in small business~\cite{Ribiero2012}. 
We combine this initial set of annotated hashtags with a larger set based on text classification.% described next.
%The set of manually curated hashtags from previous work is modest compared to the size of content disseminated in a large system like Twitter. Hence, sparsity of each hashtag usage is a limiting factor in characterizing topic-specific behavior. 

To increase the number of considered hastags, we adopt a systematic approach for annotating more hashtags based on urls within the tweets. We pair non-annotated hashtags with web urls, based on co-occurrence within posts. We extract relevant text content from each url destination (most commonly news articles from foxnews.com, cnn.com, bbc.co.uk) and build a corpus of texts related to each hashtag. We then classify the url texts in one of our 5 topics using the MALLET~\cite{McCallum2002} text classification framework trained for our topics of interest. % from two widely used topic-annotated text collections: the 20 newsgroups dataset~\cite{Rennie2008} and the News Space~\cite{Gulli}. 
As a result, we get a frequency distribution of topic classification for frequent (associated with at least $5$ texts) hashtags. The topic annotation of the hashtag is the topic of highest frequency. The number of hashtags and their usage statistics in our final topic-annotated set are presented in Table~\ref{tab:hastag_topic} (columns Users and Uses/HT). The Celebrities hashtags do not occur frequently enough in the CRAWL dataset and hence we exclude them from our analysis.

\section{Genotype model validation in Twitter}
To qualify the genotype model as a meaningful representation of social network users, we demonstrate that the genotype model is capable of capturing stable individual user behavior for a given topic. We seek to evaluate the stability of configuration of multiple users' genotype values within a topic, and use a classification task and the obtained (training/testing) accuracy as a measure of consistency for our genotype model. Within this context, we compare different genotype dimensions and evaluate the level to which each of them captures characteristic invariant properties of a social media user.

%\begin{table}
%\centering
%{\scriptsize
%\caption{Error rates for the LD local classifier algorithm for each metric and topic. The \textit{random error} describes the error rate when randomly assigning hashtags into topics.}
%\centering
%\begin{tabular}{|l|c|c|c|c|c||c|} 
%\hline                     
%	& Bus. & Celeb. & Poli. & Sci./Tech. & Sport & E[x]\\ [0.5ex]
%\hline
%Random Error	& 0.96 & 0.95 & 0.28 & 0.85 & 0.95 & 0.80 \\ \hline
%F-PAR 			& 0.30 & 0.40 & 0.54 & 0.43 & 0.31 & 0.49 \\\hline
%LAT 			& 0.29 & 0.40 & 0.44 & 0.42 & 0.41 & 0.43 \\\hline
%LOG-LAT 		& 0.40 & 0.23 & 0.40 & 0.34 & 0.38 & 0.38 \\\hline
%N-PAR 			& 0.30 & 0.39 & 0.54 & 0.43 & 0.31 & 0.48 \\\hline
%N-USES 			& 0.46 & 0.41 & 0.61 & 0.45 & 0.51 & 0.54 \\\hline
%TIME 			& 0.51 & 0.28 & 0.46 & 0.48 & 0.48 & 0.47 \\
%\hline
%\end{tabular} \label{Table:snap_local_performance}
%}
%\end{table}

\subsection{Topic consistency for individual users}
Our hypothesis is that individual users exhibit consistent behavior of adopting and using hashtags (stable genotype) within a known topic. If we are able to capture such invariant user characteristics in our genotype metrics then we can turn to employing the genotypes for applications. 
\begin{figure}[t]
\centering
 \includegraphics[width=0.4\textwidth] {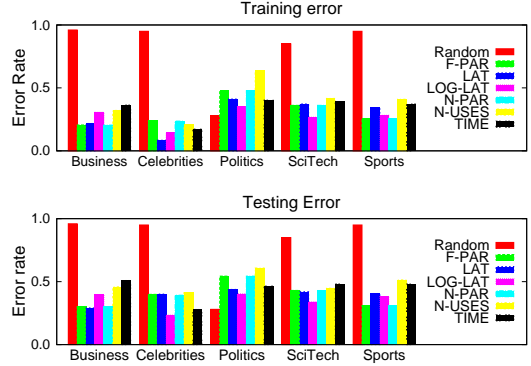}  
\caption{Training and testing accuracy of hashtag classification in a leave-one-out Linear Discriminant classification.}
\label{fig:SNAP_traintest}
\vsa
\end{figure}
We compute genotype values according to our collection of hashtags with known topics by training a per-user Linear Discriminant (LD) topic classifier to learn the separation among topics. Consider, for example, the LOG-LAT genotype metric: for a user $u$, we have a set of observed LOG-LAT values (based on multiple hashtags) that are associated with the corresponding topics. If the user $u$ is consistent in her reaction to all topics, then the LOG-LAT values per topic will allow the construction of a classifier with low training and testing error. It is also noted that each hashtag does end up having a topic distribution, but for the scope of this study, a sufficient hashtag classification should at least agree in the topic of greatest probability/likelihood, which is what is presented here.

%The LD algorithm uses a linear function to partition the hashtag topic values for each user, and this function is optimized to produce the largest possible separation between the input topic data values~\cite{Krzanowski1996}. 
The consistency of user responses is evaluated using a leave-one-hashtag-out validation: given the full set of $(u,h)$ response values, we withhold all pairs including a validation hashtag $h$ and employ the rest of the pairs involving hashtags of known topic to estimate the individual user's topic genotype. We repeat this for all genotype metrics. The training and testing error rate for this experiment are presented in Fig.~\ref{fig:SNAP_traintest}, and their similar error rates demonstrate how consistent users are at classifying hashtags into topics. In both cases, our genotype metrics enable significantly lower error rates as compared to a Random model (i.e. random prediction based on number of hashtags within a topic), demonstrating that, in general, genotype metrics capture consistent topic-wise behavior. One exception is the Politics topic as it has comparatively many more hashtags than other topics, skewing the random topic distribution resulting in slightly lower error. Across genotype metrics, we observe that normalized latency of adoption (LOG-LAT) is more consistent per user than alternatives.

\begin{table}[t]
\centering
{\scriptsize
\begin{tabular}{|l|c|c|c|c|c||c|} 
\hline                     
	& Bus. & Celeb. & Poli. & Sci./Tech. & Sport & E[x]\\ [0.5ex]
\hline
Random Error	& 0.96 & 0.95 & 0.28 & 0.85 & 0.95 & 0.45 \\ \hline
F-PAR 			& 0.50 & 0.88 & 0.61 & 0.15 & 0.09 & 0.41 \\\hline
LAT 			& 0.09 & 0.46 & 0.18 & 0.19 & 0.25 & 0.21 \\\hline
LOG-LAT 		& 0.05 & 0.13 & 0.19 & 0.12 & 0.03 & 0.13 \\\hline
N-PAR 			& 0.09 & 0.50 & 0.88 & 0.09 & 0.03 & 0.40 \\\hline
N-USES 			& 0.45 & 0.42 & 0.90 & 0.22 & 0.56 & 0.54 \\\hline
TIME 			& 1.0	& 1.0 & 0.01 & 0.92 & 0.88 & 0.61 \\\hline
\end{tabular} 
}
\caption{Error rates of the NB consensus topic classification. $E[x]$ is the expected error across topics.}
\label{tab:snap_performance}
\vsa
\end{table}

\subsection{Topic consistency within the network}
%In order to track topics, or recommend relevant content, it is essential to understand the topic of newly-arising hashtags. To this end, we leverage the existing genotypes for the SNAP dataset and build a concensus classification framework based on how new hashtags spread within the network of genotype-annotated nodes (i.e., Twitter users). We begin by using the individual user classifications from the validation set of hashtags, and then implement a Naive Bayes (NB) algorithm to achieve consensus on the topic classification of each validation hashtag. Additionally, we also demonstrate that consensus becomes more accurate as more individuals use the given hashtag.
While individual users may exhibit some inconsistencies in how they behave w.r.t. hashtags within a topic, an ensemble of users' genotypes remain more consistent overall. To demonstrate this effect, we extend our classification-based evaluation to the the network level. We implement a network-wide ensemble-based Naive Bayes (NB) classifier that combines output of individual user classifiers to achieve network-wide consensus on the topic classification of each validation hashtag. Details on the network level classifier are available in the Supplement~\cite{appendix}.

Table~\ref{tab:snap_performance} summarizes the testing error rate of our NB scheme for classifying hashtags into topics in a leave-one-hashtag-out validation. The consensus error rate decreases compared to local classifiers (Fig.~\ref{fig:SNAP_traintest}), demonstrating that the genotypes, as a complex, are more stable and consistent than individual users. The lowest error rate of $0.13$ is achieved when using the LOG-LAT metric. %The accuracy of the latency-based genotype dimensions LAT and LOG-LAT is interesting to note, because they both gauge user response times from a simple counting of Twitter posts in a certain period of time. Nonetheless, LAT and LOG-LAT are the metrics that yield the most consistent genotypes. %Based on the NB classifier performance, one can infer that users' reaction to new posts within a topic remains stable. This is true to a lesser extent for other genotype dimensions such as number of adopting parents and number of uses.

The latency genotype metrics that are most invariant (LAT and LOG-LAT) implicitly normalize their time scales of response with respect to the user's own frequency of activity, which is a feature not captured by the absolute TIME metric, or any of the other metrics. Furthermore, both of these metrics incorporate the network structure, measuring the message offset since the earliest exposure to the hashtag via a followee. LOG-LAT has a slight advantage over LAT because it suppresses the background noise of each hashtag measurement. However, LOG-LAT has the disadvantage that it is dependent on a network-wide latency measurement for the same hashtag, which might be harder to obtain in practice. In this sense, LAT is a more practical genotype dimension when summarizing individual user behavior in real time. % and employing the latter to classify new information.

\begin{figure}[t]
\centering
\subfloat[LAT Net Classifier]{\label{Fig:SNAP_all_latency_LD_NB_logfit}
 \includegraphics[width=0.23\textwidth] {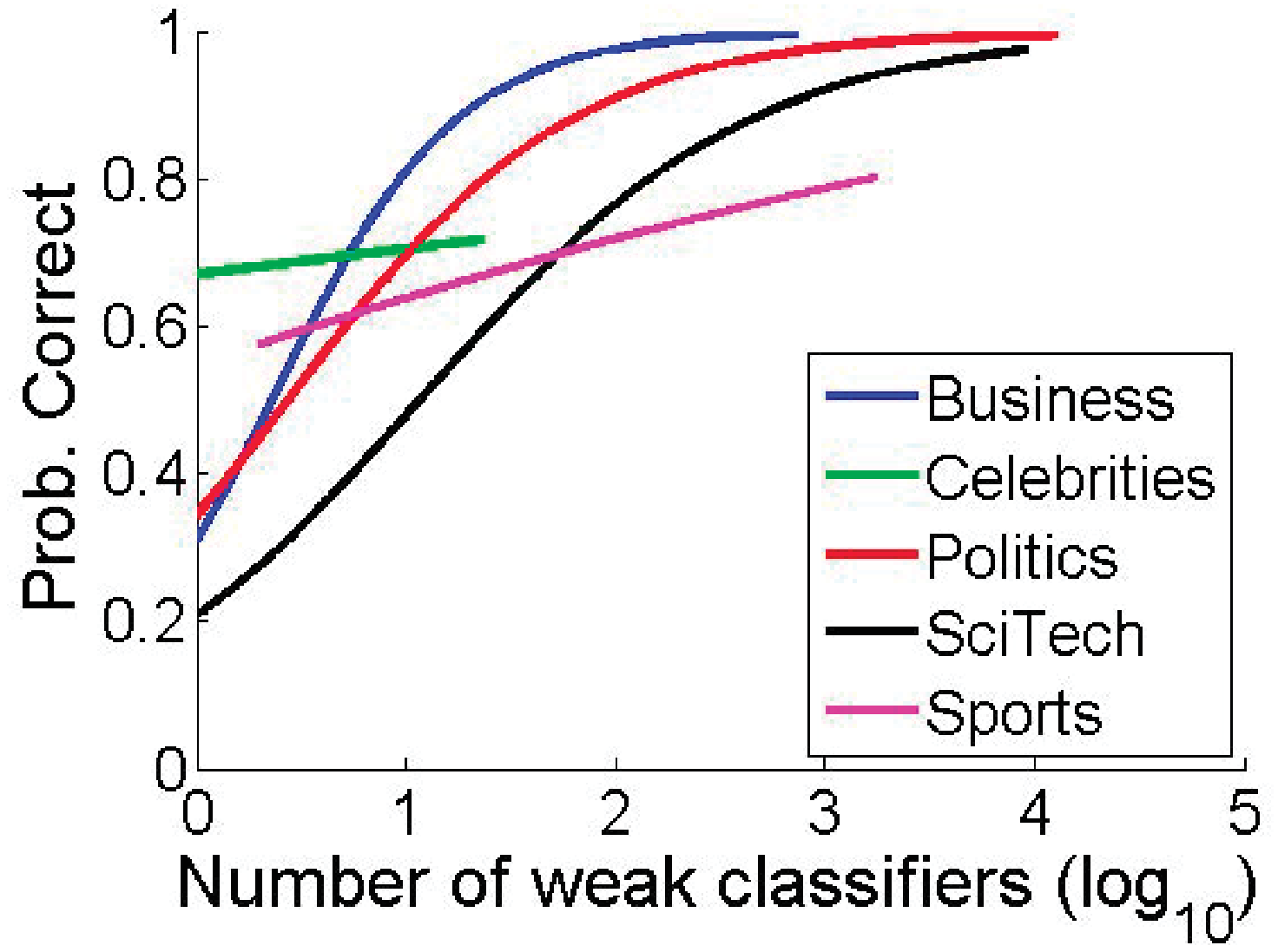}}  
\subfloat[LOG-LAT Net Classifier]{\label{Fig:SNAP_all_loglat_LD_NB_logfit}
 \includegraphics[width=0.23\textwidth] {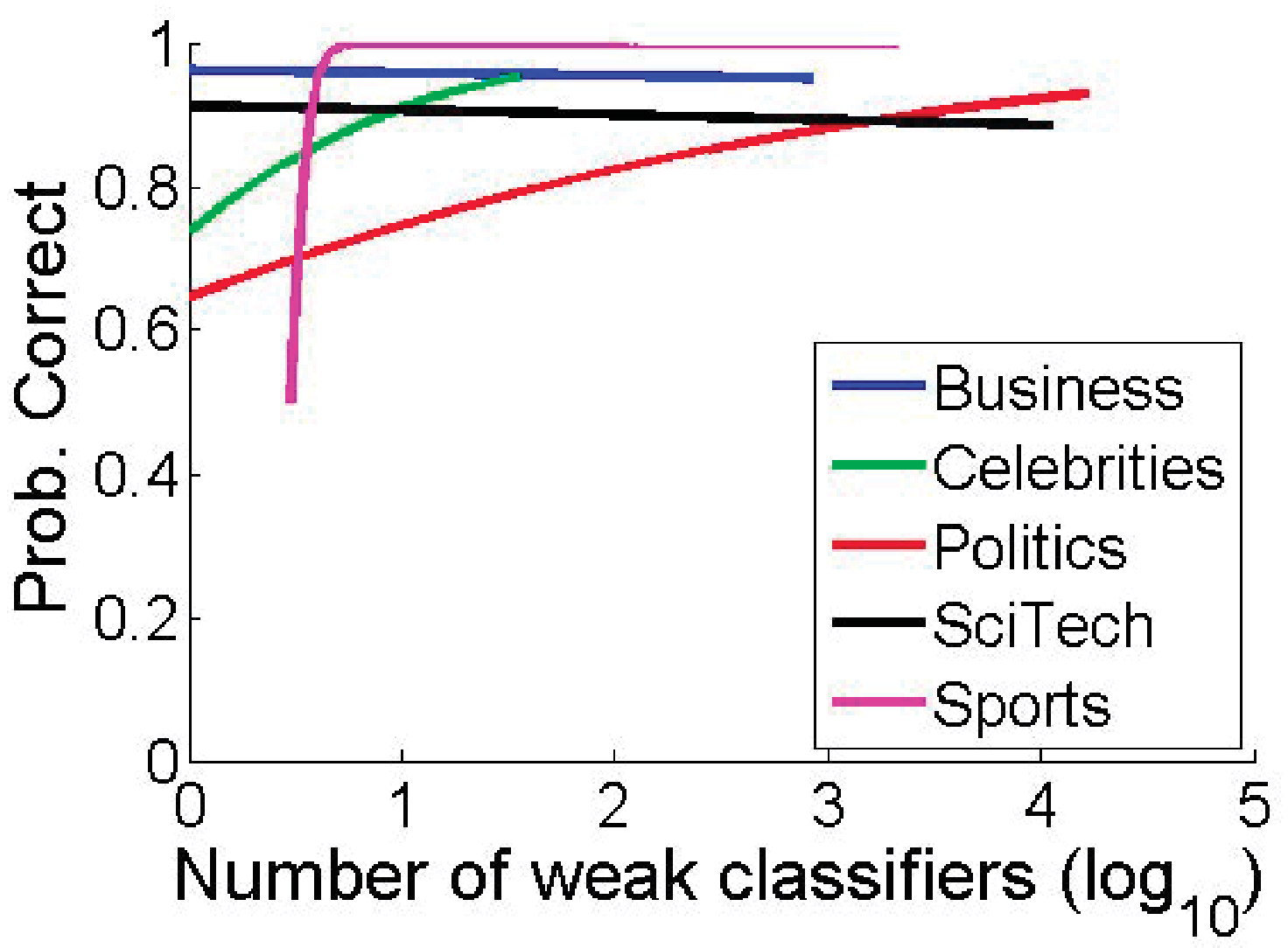}}
\caption{Accuracy of the network classification as a function of the number of local classifiers (SNAP). Logistic function is fit to each topics accuracy.}
\label{Fig:SNAP_local} 
\vsa
\end{figure}
While the system of all user genotypes exhibits significant consistency (high classification accuracy), it is interesting to understand how many user genotypes are needed to obtain a good classification (i.e. detect a network-wide topic-specific spread). We observe an increasing classification accuracy with the number of users included in the NB scheme. Figures~\ref{Fig:SNAP_all_latency_LD_NB_logfit} and \ref{Fig:SNAP_all_loglat_LD_NB_logfit} show the dependence of accuracy on number of local LD classifiers included per topic. All curves increase sharply, indicating that variability within individuals is easily overcome by considering a small subset of users within the network. The accuracy of LOG-LAT increases ``faster'' to its optimal level for increasing number of local classifiers, since the LOG-LAT metric features a network wide normalization and thus contains global information.
% The most significant consequence of our results in this section is that topic designations for hashtags can be deduced based only on the consistency of user reactions to content produced by their followees in the network, as captured by their genotype. Moreover, the accuracy improves with the number of known genotypes producing accurate topic predictions with as few as $100$ user genotypes in a large ($20k$ users) network in our experiments.

\section{Topic-specific influence backbones}
As we demonstrate in the previous section, user behavior remains consistent within a topic. A natural question inspired by this observation is whether topics propagate within similar regions of the shared medium that is the follower network structure. By observing the behavior of agents (adoption, reposting, etc.) one can reveal the underlying backbones along which topic-specific information is disseminated. In this section, we study the propagation of hashtags within Twitter to identify \textit{topical influence backbones} --- sub-networks that correspond to the dynamic user behavior. We superimpose the latter over the static follower structure and perform a thorough comparative analysis to understand their differences. The topical backbones in combination with the individual user genotypes will then enable various applications as we show in the subsequent section.     

\begin{figure}[t]
  \centering
  \includegraphics[width=0.35\textwidth]{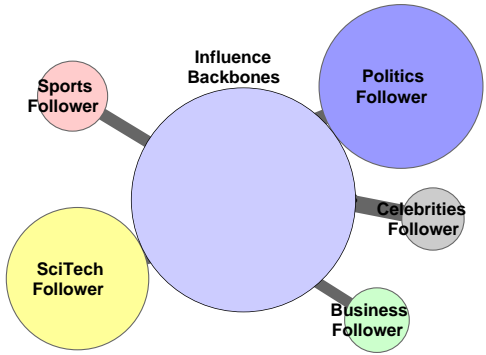} 
  \caption{Overlap among topic influence and corresponding follower subnetworks (in SNAP). Each network is represented as a node, with every topic represented by an influence (encircled in the middle) and a follower network. Node sizes are proportional to the size of the network (ranging from $120k$ for Celebrities to $42m$ for Politics Follower). Edge width is proportional to the Jaccard similarity of the networks (ranging from $10^{-3}$ inter-topic edges to $10^{-1}$ between corresponding influence-follower networks).}\label{fig:net_overlap_snap}
\vsa
\end{figure}

%% definition of influence backbone
An \textit{influence edge} $e_{i}(u,v)$ connects a followee $u$ who has adopted at least one hashtag $h$ within a topic $T_i$ before the corresponding follower $v$. Hence, the influence network $N_{i}(U,E_{i})$ for topic $T_i$ is a subnetwork of the follower network $N(U,E)$ (including the same set of nodes $U$ and a subset of the follower edges $E_{i}\in E$). We weight the edges of the influence network by the number of hashtags adopted by the followee after the corresponding follower within the same topic. 

First, we seek to understand the differences between the influence backbones and the static follower network. Figure~\ref{fig:net_overlap_snap} presents the overlap among influence backbones and their corresponding follower network. For this comparison, we augment an influence network with all follower edges among the same nodes to obtain the corresponding follower network. %\todo{By placing the definition for Jaccard Similarity in the figure caption, the next two sentences can be omitted because of redundancy.} 
In the figure, each network is represented by a node whose size is proportional to the network size (in edges). Connection width is proportional to the \textit{Jaccard Similarity (JS)} (measured as the relative overlap $|E_i\bigcap E_j|/|E_i \bigcup E_j|$) of the edge sets of the networks. The Jaccard similarity for influence and follower networks varies between $0.16$ for Sports to $0.3$ for Celebrities. The influence networks across topics do not have high overlap (JS values not exceeding $0.01$), with the exception of Sci/Tech and Politics with $JS=0.07$. This may be explained partially by the fact that these are the largest influence networks ($5$ and $11$ million edges respectively). Another reason could be that there are some ``expert'' nodes who are influential and active in both topics. The degree distribution within the topic backbones also changes (w.r.t. the static structure) with highest impact on low degree nodes and significantly lower fraction of reciprocal links between users (see Supplement~\cite{appendix}).  

\begin{figure}[t]
  \centering
  \includegraphics[width=0.42\textwidth]{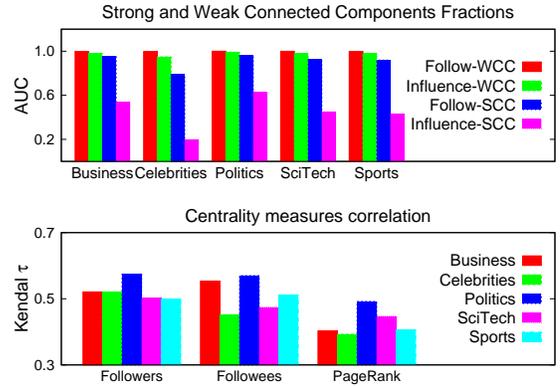}
  \caption{Largest weakly and strongly connected component (WCC and SCC) sizes as a fraction of the network size (top); and Kendall $\tau$ rank correlation of node importance measures for the influence and follower networks (bottom) (SNAP). }\label{fig:cc_snap}
\vsa
\end{figure}

%% SCC and WCC
Beyond network sizes and overlap, we also quantify the structural differences of the influence backbone in terms of connected components. A \textit{strongly connected component (SCC)} is a set of nodes with directed paths among every pair, while in a \textit{weakly connected component (WCC)} connectivity via edges regardless of their direction is sufficient. Figure~\ref{fig:cc_snap} compares the sizes of the largest SCC and WCC in the topic-specific networks as a fraction of the whole network size. When ignoring the direction (i.e. considering WCC), both the influence and follower structures have a single large component amounting to about $99\%$ of the network. The communities that are active within a topic are connected, showing a network effect in the spread of hashtags, as opposed to multiple disjoint groups which would suggest a more network-agnostic adoption. When, however, one takes direction into consideration (SCC bars in Fig.~\ref{fig:cc_snap}), the size of the SCC reduces drastically in the influence backbones. Less directed cycles remain in the influence backbone, resulting in a structure that is close to a directed acyclic graph with designated root sources (first adopters), middlemen (transmitters) and leaf consumers. The reduction in the size of the SCC is most drastic in the Celebrities topic, indicative of a more explicit traditional media structure: sources (celebrity outlets or profiles) with a large audience of followers and lacking feedback or cyclic influence. 

%% importance
\textit{How does a user's importance change when comparing influence to following?} In Figure~\ref{fig:cc_snap} (bottom) we show the correlation of node ranking based on number of followers, followees and PageRank~\cite{PageRank} in the influence and follower networks. The correlation of each pair of rankings is computed according to the \textit{Kendall $\tau$} rank correlation measure. The correlation is below $0.5$ for all measures and topics. Global network importance (PageRank) is the most distorted when retaining only influence edges ($0.4$ versus $0.5$ on average), while locally nodes with many followers (or followees) tend to retain proportional degrees in the influence network. 

%% summary
Our comparative analysis of the influence and follower structure demonstrates that the influence backbone is quantitatively different from the overall follower network. The explanation for this lies in the fact that the influence backbone is based on the dynamic behavior of users (information dissemination on specific topics), while the follower structure represents the static topic-agnostic media channels among users. Not all followees tend to exert the same amount of influence over their audiences in the actual information dissemination process, giving rise to distinct topic-specific influence backbones. We obtain similar behavior in the smaller Twitter data set CRAWL (omitted due to space limitation). 

%\section{Influence backbone}
\section{Applications of genotypes and backbones}

In this section, we employ the user genotypes and the topic-specific backbones for two important applications: (i) prediction of hashtag adopters and influencers and (ii) latency minimization of topical information spread. In both applications knowledge of individual genotypes and influence backbones enables superior performance compared to the static network structure on its own.

\noindent{\bf Topic-specific influence prediction.} We employ the influence structure and the user genotypes to predict likely influencers/adopters for a hashtag. We aim to answer the following question: \textit{Which followees are likely to influence a given user to adopt a hashtag of a certain topic and analogously which followers are likely to adopt a hashtag?} This question is of paramount importance from both research and practical perspectives. On one hand, uncovering the provider-seeker influence will further our understanding of the global network information dynamics. On the other hand, the question has practical implications for social media users offering guidelines on following high-utility sources or keeping the follower audience engaged. 

In this experiment, we consider \textit{(h,u)} pairs, for users who have at least $10$ followees and have used the hashtag at least once. The goal is to predict the subset of all followees who have used the hashtag prior to the user in question and similarly all adopting followers who are likely to use the hashtag later. 
We construct three predictors utilizing the follower structure that rank influencers/adopters by \textit{Followees}, \textit{Followers} and \textit{Reciprocal} links. We also construct three activity-based predictors utilizing genotypes and influence edges and ranking influencers/adopters by their activity \textit{Act}, topic-specific activity \textit{Topic Act} and activity combined with centrality in the corresponding backbone \textit{RW+Act}. All predictors do not have information about the spread of the specific hashtag. More details on the predictors are available in the Suppplement~\cite{appendix}.   

\begin{figure}[t]
  \centering
    \includegraphics[width=0.48\textwidth]{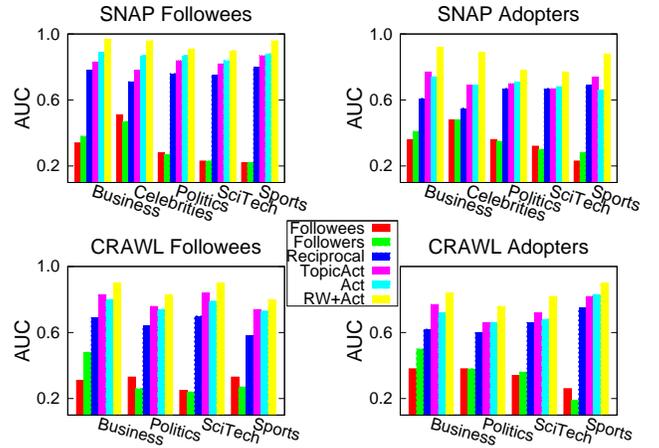} 
  \caption{Influential followee and adopter prediction accuracy. We consider several predictors of a user's influencers by a hashtag in a known topic. Genotype-based predictors (Act, Topic Act and RW+Act) perform better than follower structure-only counterparts (Follower, Followee and Reciprocal).}\label{fig:influencers}
\vsa
\end{figure}

A prediction instance is defined by a user $u$ and an adopted hashtag $h$. Only a subset $I(u,h)$ of all structural followees/followers of the user are true influencers/adopters (positives for the prediction task). Our goal is to predict the subset of true influence neighbors using their features and local influence structure (excluding information about the same hashtag $h$). A good predictor ranks the true neighbors first. In order to overcome the effect of sparsity in the data, we consider prediction of instances for which at least one candidate followee is not isolated in the influence network after removing the links associated with the target hashtag. We measure true positive and false positive rates for increasing value of $k$ (the maximal rank of predicted influencers/adopters) and compute the average area under the curve (AUC) as a measure of the predictor quality. We report this measure within each topic in Figure~\ref{fig:influencers} for the SNAP (top) and CRAWL (bottom) datasets and for influence followees (left) and adopter (right) prediction. 
% discussion of results
%% structural
Overall, in both datasets, the genotype-based predictors outperform the structure-only counterparts. The existence of a reciprocal follow link is the best structure-only predictor implying the importance of bi-directional links which often may correspond to a friendship relationship~\cite{Romero2011}. %Social friends have been found to re-share the same information with a very low latency in a recent large scale field experiment~\cite{Bakshy2012}, which may also be related to reciprocal links performing closer to the genotype-based predictors as compared to number of followees or followers.
%% behavioral
The genotype-based predictors relying on topic specific activity, overall activity and the influence structure allow over $20\%$ improvement with respect to the reciprocity predictor and above two-fold improvement compared to number of followees/followers predictors. Although node information alone (\textit{Act} and \textit{Topic Act}) provides a good accuracy, this effect is even stronger when combining them with the knowledge of the topic influence network in the composite \textit{RW+Act} predictor. The RW+Act increases the rank of followees who have influenced the same user or other users within the same topic for different hashtags.

%% crawl
The predictor performance is similar in the CRAWL dataset (Fig.~\ref{fig:influencers}), showing the generalization of our models to different types of data. The smaller improvement in CRAWL (compared to SNAP) can be explained partially by sparser usage of our analysis hashtags or due to possibly evolving genotypes of users over longer time frames, a hypothesis we are planning to evaluate as future work. %We plan to investigate the possible dynamics of genotypes in future research.
%\todo{Can one do a similar prediction in the other direction? If I know that someone tweets a hashtag, which children are going to adopt it?}
%\todo{Can one answer the question: Adoption is because of topic-influential followees as opposed to "critical mass" (number of parents who already adopted)? Note: critical mass is related to the threshold model.}

\noindent{\bf Network latency minimization.} Another important problem that can be addressed given knowledge of topic-specific user behavior is that of improving the speed of information dissemination. Fast information dissemination is critical for social-media-aided disaster relief, large social movement coordination (such as the Arab Spring of 2010), as well as time-critical health information distribution in developing regions. In such scenarios, genotypes and the influence structure among users are critical for improving the overall ``latency'' of the social media network. In this subsection, we demonstrate the utility of our individual user models for latency minimization. 

%% Define the problem
Consider a directed path in a topic-specific influence backbone $N$, defined by a sequence of nodes $P=(u_1,u_2...u_k)$. The \textit{path latency} $l(P)$ is defined as the sum of topic-specific latencies (\textit{Time} measure of the genotype) $l(P)=\sum_{j=1...k-1}Time(u_j),u_j \in P$ of all nodes except the destination. The \textit{source-destination latency} (or just latency) $l(u_1,u_k)=\min_{P:u_1\rightarrow u_k}l(P)$ is defined as the minimum path latency considering all directed paths between the target nodes. The concept of latency is similar to that of shortest path length, except that ``length'' is measured according to the responsiveness of traversed nodes (i.e., minimal time until $u_k$'s adoption of a hashtag introduced by $u_1$). The \textit{average network latency} is defined as the mean of all node pair latencies.
\begin{figure}[t]
  \label{fig:minlat}
  \centering
  \includegraphics[width=0.4\textwidth]{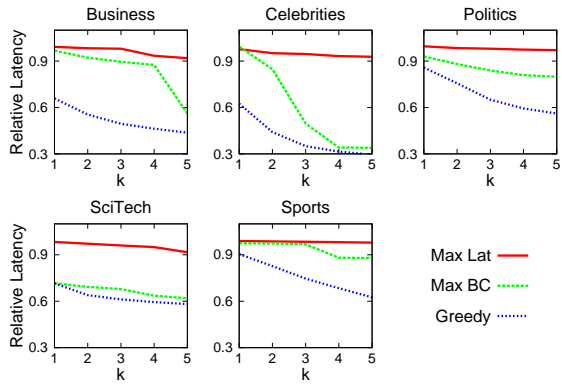} 
  \caption{Comparison of three heuristics for Latency Minimization in the SNAP dataset. The traces show the relative (w.r.t. the original) average network latency as a function of the number of targeted nodes $k$.}\label{fig:minlat}
\vsa
\end{figure}
%% Latency minimization
Given a directed network $N(V,E)$ and latency for every node, we define the problem of \textit{$k$ Latency Minimization ($k$-LatMin)} as finding the $k$ best target nodes, whose latency reduction leads to the largest average network latency decrease. We assume that specific nodes could be targeted to reduce their individual latency. In real application scenarios, node latency can be reduced by timely and relevant content recommendation to target nodes and/or financial incentives. For our analysis we optimistically assume that every node's latency could be reduced to $0$, however, node-wise constraints can be incorporated according to known limitations of users. 

One can show (via a reduction from the \textit{Set cover} problem) that $k$-LatMin is NP-hard. %We leave the complexity analysis as well as possible sub-modularity investigation of the average latency function as future work. 
% The reduction intuition is as follows: from a Set cover instance with $m$ elements and $n$ subset and cover size $k$, we construct a bipartite graph with $m$ element nodes of zero latency nodes and $n$ set nodes if latency $1$. Set and element nodes are connected if the element is contained in the corresponding set. We require that we obtain $k$-MinLat solution with average latency less than $\frac{(n-k)}{(m+n)}$, i.e. by obtaining a k-cover the corresponding targeted nodes will incur latency cost of $1$ for only paths starting in non-covered set nodes.
%% explain heuristics
We consider three heuristics: \textit{Max Lat} targets nodes in descending order of their latency values; \textit{Max BC} targets nodes in decreasing order of their structural node betweeness-centrality measure; and \textit{Greedy} targets nodes based on their maximal decrease of average latency combining both structural (centrality) and genotype (latency) information. 

Figure~\ref{fig:minlat} shows the performance of the three heuristics in minimizing the average latency in subgraphs (of size 500 nodes) of the largest strongly connected components within the influence backbones of our SNAP dataset. Considering the node genotypes (Max Lat) or the influence backbone (Max BC) on their own is less effective than jointly employing both (Greedy) across all topics. The Greedy heuristic enables about 2-fold reduction of the overall network latency by targeting as few as 1\% (5 out of 500 nodes) of the user population. It is interesting to note that in Sports and Celebrities, since there are central nodes of large degrees, the betweeness-centrality criterion performs almost as good as Greedy.   
%% re-iterate that this is just a demonstration of the utility of genotypes issues   

\balance

\section{Conclusion}
We introduced the social media genotype---a genetically-inspired framework for modeling user participation in social media. Features captured by the user genotypes define the \textit{actual topic-specific} user behavior in the network, while the traditionally analyzed follower network defines only \textit{what is possible} in the information dissemination process. Within our genotype model, each network user becomes an individual with a unique and invariant behavioral signature within the topic-specific content dissemination. In addition, we demonstrated that users are embedded in topic-specific influence backbones that differ structurally from the follower network.

We instantiated our topic-based genotype and backbone framework within a large real-world network of Twitter and employed it for the tasks of \textit{(i)}  discovering topic-specific influencers and adopters, and \textit{(ii)} minimizing the network-wide information dissemination latency. The genotype framework, when combined with the topic-specific influence backbones, enabled good influence predictive power, achieving improvement by more than $20\%$ over using the follower structure alone. In the latency minimization application, we demonstrated that the knowledge of topic backbones and genotypes can enable 2-fold reduction of the overall network latency by reducing the latency of appropriately selected 1\% of the user population.
 
\noindent{\bf Acknowledgements.} This work was supported by the Institute for Collaborative Biotechnologies through grant W911NF-09-0001 from the U.S. Army Research Office and by the Army Research Laboratory under cooperative agreement W911NF-09-2-0053 (NS-CTA). The content of the information does not necessarily reflect the position or the policy of the Government, and no official endorsement should be inferred. The U.S. Government is authorized to reproduce and distribute reprints for Government purposes notwithstanding any copyright notice herein.

{\footnotesize

\end{document}